\documentstyle[12pt]{article}
\def\beq{\begin{equation}}
\def\brr{\begin{array}}
\def\err{\end{array}}
\def\eeq{\end{equation}}
\def\bea{\begin{eqnarray}}
\def\eea{\end{eqnarray}}

\def\ni{\noindent}

\def\nn{\nonumber}
\def\ms{\medskip}

\begin{document}

\hfill UB-ECM-PF 92/12
\mbox{}

\vspace*{1cm}

\begin{center}

{\LARGE \bf
Higher derivative quantum gravity with torsion in the conformally
self-dual limit}

\vspace{8mm}

{\sc E. Elizalde and S.D. Odintsov}\footnote{On leave from
Department of
Mathematics and Physics,
 Pedagogical Institute, 634041 Tomsk, Russia.}
\ms

Department E.C.M., Faculty of Physics, \\
University of Barcelona, \\
Diagonal 647, 08028 Barcelona, Spain \\
{\it e-mail: eli @ ebubecm1.bitnet}
\vspace{1cm}

{\sl May 11th, 1992}

\vspace{1cm}

{\bf Abstract}

\end{center}

The path integral for higher-derivative quantum gravity with
torsion is considered. Applying the methods of two-dimensional
quantum gravity, this path integral is analyzed in the limit of
conformally self-dual metrics. A scaling law for fixed-volume
geometry is obtained.

\vspace{1cm}

\noindent PACS: \begin{quote} 03.70 Theory of quantized fields, \
04.50 Unified theories and other theories of gravitation, \
11.10 Field theory.
\end{quote}

\newpage

It is widely known that higher-derivative quantum gravity is
multiplicatively renormalized [1] and that it is an asymptotically
free theory [2]. It provides a very natural framework for a
consistent theory of quantum gravity and for unification with
matter. Unfortunately, in spite of the numerous efforts by
different people [1-5], the unitarity of this theory  is still an
unsolved problem (for a general review of higher-derivative
gravity and of GUTs interacting with higher-derivative gravity
see [6]).

 On the other side, it is known that higher-derivative quantum
gravity with torsion can also be multiplicatively renormalized,
and that it is unitary (free of ghost poles), at least for some
values of the parameters involved an at the tree level [7]. It
can be therefore interesting to study the properties of this
theory on the quantum level in some detail, because it can
provide a new framework for a consistent theory of quantum
gravity. However, higher-derivative gravity with torsion  [7] is
very complicated ---already at the classical level--- and some
reasonable simplifications must be introduced from the begining.

A slightly simplified version of the full theory, which has been
discussed in [6], is described by the following Lagrangian:
\bea
L_G &=& \lambda_0 + \gamma_0 R + \eta_0 R^2 + \rho_0 F +
\alpha_{01} F^2_{\mu \nu}+  \alpha_{02} (\nabla_{\mu} S^{\mu})^2+
\alpha_{03} (S_{\mu} S^{\mu})^2 \nn \\
&+& \alpha_{04} R^{\mu \nu} S_{\mu} S_{\nu} + \alpha_{05} R
S_{\mu} S^{\mu} + \gamma_0 \alpha_{06}  S_{\mu} S^{\mu} +
\mbox{surface terms}.
\eea
Here $F=C^2_{\mu\nu\alpha\beta}$ is the Weyl tensor, $R_{\mu\nu}$
the Ricci tensor, and $R$ the curvature, associated with
$\nabla_{\mu}$ (the covariant derivative without torsion),
$F_{\mu\nu} =\nabla_{\mu} S_{\nu} -\nabla_{\nu} S_{\mu} $, with
$S^{\nu} = i \epsilon^{\alpha \beta\mu\nu} T_{\alpha\beta\mu}$,
being $ T_{\alpha\beta\mu}$ the torsion tensor and, finally,
$\lambda_0, \ldots, \alpha_{06}$,  are the bare coupling
constants of the theory.

We shall consider the case when only the antisymmetric part of
the torsion (i.e., $S_{\mu}$) is non-zero, while all the other
components of the torsion are zero. Notice that torsion can
minimally interact with matter (spinors) only through the
$S_{\mu}$. Another reason in favor of our restriction is the fact
that the vector $S_{\mu}$ is most important for the cosmological
applications of gravity with torsion.

A theory with the Lagrangian (1) is multiplicatively
renormalized. The perturbative approach to this theory has not
been developed yet (one may even question its existence), due to the
presence of the vector
field $S_{\mu}$, which is {\sl not} the gauge field. We do not have a
gauge-fixing term for it, and we do not even know how to
construct the propagator.

The purpose of this letter  is to show that some quantum elements
and properties  of the theory (1) ---such as its path integral in
the self-dual limit and some scaling laws--- can be investigated.
The path integral which is interesting for our analysis is the
following (written in Euclidean coordinates)
\beq
\int {\cal D} g_{\mu\nu}  {\cal D} S_{\alpha}  {\cal D} \psi \,
\exp \left( -S_m [g, S_{\mu}, \psi] - \int_{\cal M} d^4 x \, \sqrt{g}
L_G \right) O_1 [g, S_{\mu}, \psi]  \cdots O_n [g, S_{\mu},
\psi],
\eeq
where $\psi$ is the set of matter fields, $S_m$ is the conformally
invariant free matter action, $L_G$ is given by eq. (1), the
manifold ${\cal M}$ has a fixed topology, and the $O_i$, $i=1, \ldots,
n$, are some operators.

We now follow ref. [8], where the path integral (2) without
torsion  and in the       self-dual limit   $\rho_0 \rightarrow
\infty$ has been studied. Presumably, such  limit describes the
infrared dynamics of four-dimensional quantum gravity [9] (see
also [10]). When   $\rho_0 \rightarrow \infty$, the path integral
over $g_{\mu\nu}$ becomes the integral over conformally self-dual
metrics. Notice that the delta function arising from the conformally
self-dual part of the Weyl tensor reduces the integral over the 10
components of  $g_{\mu\nu}$ to an integral over 5 components only
[8].

Explicitly, the conformally self-dual metric has the form
\beq
g_0 = \left( \widehat{g} (m_i) e^{\Phi} \right)^{\xi},
\eeq
where $\widehat{g} (m_i)$ can be fixed via $\widehat{R}=0$ and
$\partial^{\mu} \widehat{g}_{\mu\nu}=0$. The $m_i$ parametrize the
moduli space of conformally self-dual metrics (see [11]) up to
Weyl transformations, $g_{\mu\nu} \rightarrow g_{\mu\nu}
e^{\Phi}$, and diffeomorphisms, $x^{\mu} \rightarrow x^{\mu} +
\xi^{\mu}$ (see [8] for details). With the above choice,
\beq
\delta g_{\mu\nu} =  g_{0\mu\nu}  \delta\Phi + \nabla_{(\mu}
\delta \xi_{\nu)} + \delta \bar{h}_{\mu\nu}.
\eeq
In the limit $\rho_0 \rightarrow \infty$, the integral over
$\delta \bar{h}_{\mu\nu}$ can be calculated, with the following
result [8]
\bea
&& \int S_{\mu} \int \prod_i dm_i {\cal D} \Phi \det ( O^+ O )^{-
1/2}_{\widehat{g} e^{\Phi}} \det (L^+ L)^{1/2}_{\widehat{g} e^{\Phi}} \nn
\\
&\times& \det \left( \Box^{mat} (S_{\mu}) \right)^{-1/2}_{\widehat{g}
e^{\Phi}}  \exp \left[ - \int d^4x \left. \sqrt{g} L_G \right|_{\rho=0}
\right], \eea
where the contribution of the matter partition function is
denoted by $\det \left( \Box^{mat} \right)^{-1/2}$ (of course,
for spinors the sign of the exponent, $-1/2$, is reversed), and
the number of degrees of freedom (spinors, gauge fields) should
be also taken into account. Moreover, $(L^+ L)^{\mu}_{\nu}= -2
(\delta^{\mu}_{\nu} \Box + \frac{1}{2} \nabla^{\mu} \nabla_{\nu}
+ R^{\mu}_{\nu})$. $O^+ O$ is a conformally invariant
fourth-order differential operator acting on tensors and coming
from the expansion of $C^2_{\mu\nu\alpha\beta}$ (see [8,12]).
Finally, the integral over the diffeomorphism group, ${\cal D}
\xi$ has been dropped.

Let us concentrate now on  the integral $\int {\cal D} \Phi$. To
decouple the determinants from $\Phi$ we should look to the
conformal anomaly of matter. The conformal matter action is given by
[6]\footnote{For a detailed (non-trivial) discussion of the real and
imaginary parts of the euclidean matter action in the (more simple) case
of constant background fields see [13].}
 \bea
S_m&=& \int d^4x \, \sqrt{g} \left[ \sum_i \varphi_i \left(
\Box - \frac{1}{6} R - \xi S_{\mu} S^{\mu} \right) \varphi_i
\right. \nn \\ &+& \left. \sum_j i \bar{\psi}_j \left(
\gamma^{\mu} \nabla_{\mu} - \zeta \gamma_5 \gamma^{\mu} S_{\mu} \right)
\psi_j \right].
\eea
Here $\xi$ and $\zeta$ are the constants of non-minimal coupling
of matter with torsion, and the $\varphi_i$ and $\psi_j$ are the
scalars and spinors, respectively. Note that the values $\xi=0$,
$\zeta=1/8$ correspond to the minimal interaction of matter with
torsion. (In two dimensions matter does not interact minimally
with torsion). Note also that the action of gauge fields is the
same as in the absence of torsion.

The trace anomaly has the following form [6]
\bea
<T_{\mu}^{\mu} >& =& \frac{2}{\sqrt{g}} \, \frac{\delta
S[\widehat{g}, \Phi, S_{\mu}]}{\delta \Phi} = -\frac{1}{(4\pi)^2}
\left[ a \left( F - \frac{2}{3} \Box R \right) + bG \right. \nn
\\
&+& \left. a_1 F_{\mu\nu}^2 + a_2 (S_{\mu}S^{\mu})^2 + a_3 \Box
(S_{\mu}S^{\mu}) + a_4 \nabla_{\mu} (S_{\nu} \nabla^{\nu} S_{\mu}
- S^{\mu} \nabla_{\nu} S^{\nu}) \right] \\
&+& \lambda' + \gamma' R + \gamma' {\alpha'}_6 S_{\mu}S^{\mu} +
{\alpha'}_2 (\nabla_{\mu}S^{\mu})^2 +{\alpha'}_4 R^{\mu\nu}
S_{\mu}S_{\nu} + {\alpha'}_5 R S_{\mu}S^{\mu} + b' \Box R, \nn
\eea
where the values (finite) of $a$ and $b$ are well known (see, for
example, [12]), and [6]
\beq
a_1= a_4= - \frac{2}{3} \zeta^2, \ \ \ a_2= \frac{1}{2} \xi^2, \
\ \  a_3= \frac{1}{3} \left( 2\zeta^2-  \frac{1}{2} \xi^2\right).
\eeq
The divergent parameters $\alpha', \gamma', \ldots, b'$,
 renormalize the corresponding
coupling constants.

Using the equation which is valid for conformally invariant
differential operators
\beq
\det X_{\widehat{g} e^{\Phi}} = \det X_{\widehat{g}}  \exp \left( -
S[\widehat{g}, \Phi, S_{\mu}] \right),
\eeq
where $ X_{\widehat{g}}$ can depend on $S_{\mu}$ ($\widehat{S}_{\mu} =
S_{\mu}$), if $ X_{\widehat{g} e^{\Phi}}$ depends on  $S_{\mu}$, and
integrating over the trace anomaly (7) [14] (see also [15,9]) for
the case $S_{\mu}=0$), we can rewrite (5) as follows
\beq
\int {\cal D} S_{\mu} \int \prod_i dm_i \chi (m_i, S_{\mu}) \int
{\cal D} \Phi \,  \exp \left( -S[\widehat{g}, \Phi, S_{\mu}] \right),
\eeq
where
\beq
\chi (m_i, S_{\mu}) = \det ( O^+ O )^{-
1/2}_{\widehat{g} (m_i)} \det (L^+ L)^{1/2}_{\widehat{g} (m_i)}
 \det \left( \Box^{mat} (S_{\mu}) \right)^{-1/2}_{\widehat{g} (m_i)}
\eeq
depends also on the torsion $S_{\mu}$,
 due to the interaction of matter with torsion.
We have
\bea
S[\widehat{g}, \Phi, S_{\mu}]  &=& \frac{B_0}{32\pi^2}  S_0[\widehat{g},
\Phi] + \frac{A_0}{32\pi^2}  S_1[\widehat{g}, \Phi] + \eta_1 S_{R^2}
\nn \\
&+& \gamma_1 S_{R} + \lambda_1 S_{cc} + \left( \alpha_1 +
\frac{a_1}{(4\pi)^2} \right) S_F +\left( \alpha_3 +
\frac{a_2}{(4\pi)^2} \right) S_4 \\
& +&  \frac{1}{(4\pi)^2} \left( a_3 + \frac{1}{2} a_4 \right) S_2
+ \frac{a_4}{(4\pi)^2} S_{21} + \cdots \nn
\eea
Here
\bea
&& A_0= a_0+a_L+a_{mat}, \ \ \ \   B_0= b_0+b_L+b_{mat},  \nn \\
&&   S_0[\widehat{g}, \Phi] = \int d^4x \, \sqrt{\widehat{g}} \left[ \Phi
\Delta \Phi + \left( \widehat{G} - \frac{2}{3} \widehat{\,\Box\,} \widehat{R}
\right) \Phi \right], \nn \\
&& \Delta = \widehat{\,\Box\,}^2 + 2\widehat{R}^{\mu\nu} \widehat{\nabla}_{\mu}
\widehat{\nabla}_{\nu} - \frac{2}{3} \widehat{R} \widehat{\,\Box\,} +
\frac{1}{3???}
(\widehat{\nabla}^{\mu} \widehat{R})  \widehat{\nabla}_{\mu}, \nn \\
&&   S_1[\widehat{g}, \Phi] = \int d^4x \, \sqrt{\widehat{g}} \widehat{F}
\Phi, \nn \\
&&   S_{R^2} = \int d^4x \, \sqrt{\widehat{g}} \left[ \widehat{R} -
\frac{3}{2} (\widehat{\nabla} \Phi)^2 - 3\widehat{\,\Box\,} \Phi \right]^2,
\\
&&   S_{R} = \int d^4x \, \sqrt{\widehat{g}} \left[ \widehat{R} -
\frac{3}{2} (\widehat{\nabla} \Phi)^2 - 3\widehat{\,\Box\,} \Phi \right], \nn
\\
&&   S_{cc} = \int d^4x \, \sqrt{\widehat{g}}  e^{2\Phi}, \ \ \ \
S_F = \frac{1}{2} \int d^4x \, \sqrt{\widehat{g}}  \widehat{F}_{\mu\nu}^2
\Phi, \nn \\
&&   S_4 = \frac{1}{2} \int d^4x \, \sqrt{\widehat{g}}
\widehat{S}_{\mu}^4 \Phi, \ \ \ \  S_2 = \frac{1}{4} \int d^4x \,
\sqrt{\widehat{g}}  \widehat{S}^2 \widehat{\nabla}_{\mu} \Phi
\widehat{\nabla}^{\mu} \Phi, \nn \\
&& S_{21} = \frac{1}{4} \int d^4x \, \sqrt{\widehat{g}}
\widehat{S}^{\mu} \widehat{S}^{\nu} \widehat{\nabla}_{\mu} \Phi
\widehat{\nabla}_{\nu} \Phi. \nn
\eea
Here $\widehat{S}_{\mu}=S_{\mu}$, and $a_0, b_0, a_L$ and $b_L$ are the
contributions from $O^+O$ and $L^+L$ to the trace anomaly.

We do not write explicitly in (12) the terms dependent on torsion
in which there is no finite contribution from the conformal
anomaly. (We will be interested only in the critical point, where
the corresponding renormalized coefficients in (1), i.e.,
$\alpha_2$, $\alpha_4$,  $\alpha_5$ and  $\alpha_6$ are equal to
zero). Note also that the $\Phi$-independent terms have been
omitted in (12). It is interesting to notice that torsion in (12)
has appeared in a much more complicated way than in two dimensions,
where it can be integrated over. Here we cannot even decouple the
$\Phi$ and $S_{\mu}$ terms in the path integral.

Now, the measure ${\cal D}_{\widehat{g} e^{\Phi}} \Phi$ can be
treated as in two dimensions [16], and the $\Phi$-de\-pend\-ence can
be absorbed in the coefficients of the gravitational part of (12)
($S_{\mu}=0$), as it has been done in [8]. It changes the
coefficients $A$ and $B$, due to the contribution of the operator
$\Delta$:
\bea
A&=& \frac{1}{120} (N_0+6N_{1/2} + 12N_1 -8) + \frac{199}{30}, \\
B&=& -\frac{1}{360} (N_0+11N_{1/2} + 62N_1-28) - \frac{87}{20},
\eea
where the last term in both brackets is the contribution of
$\Delta$ to the conformal anomaly [10,12], and the last terms in
(14) and (15)
are the $a_0+a_L$ and $b_0+b_L$ found in [12], respectively. These
contributions
are equal to the trace anomaly coefficients of conformal quantum
gravity\footnote{This fact also
ensures the validity of the so-called conformal regularization
procedure [17] in conformal Weyl gravity.} (see ref. [2]).\footnote{It
would be interesting to understand whether the coefficients $A$ and $B$
(i.e., their signs) can be changed in some other conformal gravity
model. For example, for the second-order conformal gravity of [18]
(i.e., $O^+O $ is of second order), we have $b_0+b_L=-27/20$. Surely,
the signs of $A$ and $B$ can be changed in conformal supergravity
(second ref. [15]).}

The additional contributions to the terms of conformal anomaly
connected with torsion are
given only by the $\Delta$-operator (which accounts for the torsionic
terms from (13)). The direct calculation we have done using the
algorithm of fourth order operator divergencies evaluation, gives for
the coefficient which is of interest to us:
\beq
a_2^{(\Delta)} = \frac{1}{4B_0^4} \left[ \frac{1}{2} a_3^2 + \frac{3}{4}
a_3a_4 + \frac{5}{16} a_4^2 \right].
\eeq
Here, $B_0$ (15) is given by the matter part only (the first three terms
in (15)).

With $A$ and $B$ known, one can start the calculation of the
vertex operators. For example, we can look to the scaling
(classical plus anomalous) dimension of $e^{\alpha \Phi}$
corresponding to $ S_{cc} = \int d^4x \, \sqrt{\widehat{g}}
e^{2\alpha\Phi} $ and find it in terms of $B$ [8]. We can
also look to other gravitational [10] or torsion vertex operators.

Finally, we will try to define the scaling law in the presence
of torsion. Starting from the fixed-volume partition function at
the critical point (as in [8], $\eta_1=\gamma_1=\lambda_1=0$ and the
torsionic coupling constants  where there is no contribution from the
conformal anomaly are also equal to zero), we get
\bea && \int dS_{\mu} \, Z(V,S_{\mu}) = \int {\cal D} S_{\mu} \int
\prod_i dm_i \chi (m_i, S_{\mu}) \int {\cal D}_{\widehat{g}} \Phi \nn
\\ &\times&   \exp \left\{ -\frac{B}{32\pi^2} S_0[\widehat{g},
\Phi] -\frac{A}{32\pi^2} S_1[\widehat{g}, \Phi] -\frac{1}{(4\pi)^2}
\left( a_3 + \frac{1}{2} a_4 \right) S_2 [\widehat{g}, \Phi, S_{\mu}]
\right. \\
 &-& \left. \frac{a_4}{(4\pi)^2} S_{21} [\widehat{g}, \Phi, S_{\mu}]-
\left( \alpha_1 + \frac{a_1}{(4\pi)^2} \right) S_F -\left(
\alpha_3 + \frac{a_2}{(4\pi)^2} \right) S_4  \right\} \, \delta
\left(   \int d^4x \, \sqrt{\widehat{g}} e^{2\alpha\Phi} -V \right).
\nn
\eea
Using the properties of the exponential in (13) under the
constant shift $\Phi \rightarrow \Phi +c$, we get
\beq
Z(V,S_{\mu}^2 ) = \exp \left\{ \left[ -2\alpha -B \chi +
\frac{3}{2} A \tau - \left( \alpha_3+\frac{a_2}{(4\pi)^2} \right)
\bar{S}_4 \right] c \right\}
\ Z(e^{-2\alpha c} V,S_{\mu}^2 ),
\eeq
where $\chi$ is the Euler characteristic of the manifold, $\tau$
its signature, and $\bar{S}_4 =\frac{1}{2} \int d^4x \,
\sqrt{\widehat{g}} S_{\mu}^4$. It follows from (17), that
\beq
Z(V,S_{\mu}^2 ) \sim \exp \left\{ \left[ -1 + \frac{1}{4\alpha}
\left( -2B \chi + 3A \tau  -2 \left(
\alpha_3+\frac{a_2}{(4\pi)^2} \right) \bar{S}_4 \right) \right] \ln V
\right\}.
\eeq
The scaling law (18) can be the starting point for computer simulations
 of the theory. Of course, one can look to other scaling laws
corresponding to fixed torsion.

In conclusion, we have discussed here the path integral
corresponding to higher derivative quantum gravity with torsion
in the self-dual limit, generalizing the work [8]. We have found
that torsion actually shows up in a very complicated way, and
that it cannot be integrated over, as in two dimensions [19].
Torsion does not decouple from the conformal factor in this
formalism. Of course, one can guess that using some other
variables (as, for example, vielbein and connection) the
situation could be improved.

Finally, the above approach can be critized on general grounds,
because even the classification of four-dimensional manifolds is
not known (if it at all exists), and here we cannot expect that
the methods of two-dimensional quantum gravity can be very useful
in four dimensions. Nevertheless, our point is that already the
fact that these methods can be applied (as we have shown) in four
dimensions for the discussion of some limiting cases of the
theory (1) ---which probably cannot be formulated at all at the
perturbative level--- actually seems very promising.


\vspace{2cm}

\ni{\large \bf Acknowledgments}

We would like to thank Profs. Ignatios Antoniadis and Rolf Tarrach
 for fruitful discussions on related
problems.
 S.D.O. thanks the members of the Department E.C.M. of
Barcelona University for the kind hospitality.
E.E. thanks the Alexander von Humboldt Foundation for continued
logistic help.
This work has
been supported by Direcci"n Ge\-ne\-ral de Investigaci"n
Cient!fica y Tcnica (DGICYT), research projects
 PB90-0022 and SAB92-0072.


\newpage

\begin{thebibliography}{99}

\bibitem{}
K.S. Stelle, {\sl Phys. Rev.} {\bf D16} (1977) 953; {\sl Gen. Rel.
Grav.} {\bf 9} (1978) 353.

\bibitem{}
 E.S. Fradkin and A.A. Tseytlin, {\sl Nucl.
Phys.} {\bf B201} (1982) 469.

\bibitem{} A. Salam and S. Strathdee, {\sl Phys. Rev.} {\bf D18}
(1978) 4480.

\bibitem{} T. Antoniadis and E.T. Tomboulis, {\sl Phys. Rev.} {\bf D33}
(1986) 2756.

\bibitem{} S.W. Hawking, in {\sl Quantum Field Theory and Quantum
Statistics}, eds. I.A. Batalin, C.J. Isham and G.A. Vilkovisky
(Hilger, Bristol, I987).

\bibitem{} I.L. Buchbinder, S.D. Odintsov and I.L. Shapiro, {\sl
Effective Action in Quantum Gravity}
(Hilger, Bristol, 1992); see also {\sl Rivista Nuovo Cim.} {\bf 12}
(1989) 1.

\bibitem{} E. Sezgin and P. van Nieuwenhuizen, {\sl Phys. Rev.} {\bf
D21} (1980) 3269.

\bibitem{} C. Schmidhuber, preprint CALTECH, 1992.

\bibitem{} I. Antoniadis and E. Mottola, {\sl Phys. Rev.} {\bf D45}
(1992) 2013.

\bibitem{} S.D. Odintsov, {\sl Z. Phys. C}, 1992, in press.

\bibitem{} T. Eguchi, P.B. Gilkey and A.J. Hanson, {\sl Phys.
Rep. } {\bf 66} (1980) 213.

\bibitem{} I. Antoniadis, P.O. Mazur and E. Mottola, preprint LA-UR-92,
1992.

\bibitem{} E. Elizalde, {\sl Phys. Lett.} {\bf 145B} (1984) 271.

\bibitem{} I.L. Buchbinder, S.D. Odintsov and I.L. Shapiro, {\sl
Phys. Lett.} {\bf B162} (1985) 193.

\bibitem{} R.J. Reigert, {\sl Phys. Lett.} {\bf B134} (1984) 56;
E.S. Fradkin and A.A. Tseytlin, {\sl Phys. Lett.} {\bf 134} (1984) 187;
E.T. Tomboulis, {\sl Nucl. Phys.} {\bf B329} (1990) 410;
S.D. Odintsov and I.L. Shapiro, {\sl Class. Quant. Grav.} {\bf 8} (1991)
L57.

\bibitem{} V. Knizhnik, A.M. Polyakov and A.B. Zamolodchikov, {\sl Mod.
Phys. Lett.} {\bf A3} (1988) 819;
J. Distler and H. Kawai, {\sl Nucl. Phys.} {\bf B321} (1989) 509;
F. David, {\sl Mod. Phys. Lett.} {\bf A3} (1988) 1651.

\bibitem{} F. Englert, C. Truffin and R. Gastmans, {\sl Nucl. Phys.}
{\bf B117} (1976) 407; E.S. Fradkin and G.A. Vilkovisky, {\sl Phys.
Lett.} {\bf B73} (1978) 209.

\bibitem{} P. Pascual, J. Taron and R. Tarrach, {\sl Phys. Rev.} {\bf
D39} (1989) 2993.

\bibitem{} T. Berger and I. Tsutsui, {\sl Nucl. Phys.} {\bf B333}
(1990) 245.

\end{thebibliography}
\end{document}